\PassOptionsToPackage{dvipsnames}{xcolor}
\documentclass[11pt]{article}
\usepackage[utf8]{inputenc}
\usepackage{amssymb}
\usepackage{amsfonts}
\usepackage{amsmath}
\usepackage{graphicx}
\usepackage{pgfplots}
\pgfplotsset{compat=1.18}
\pdfoutput=1
\usepackage[a4paper,width=150mm,top=25mm,bottom=25mm]{geometry}
\usepackage{fancyhdr}
\usepackage{upgreek}
\usepackage{jheppub}
\newcommand{\mycomment}[1]{}

\providecommand{\U}[1]{\protect\rule{.1in}{.1in}}
\providecommand{\U}[1]{\protect\rule{.1in}{.1in}}
\definecolor{blue}{rgb}{0,0,1}

\definecolor{red}{rgb}{1,0,0}

\begin{document}
\title{Logarithmic corrections to the entropy of near-extremal black holes in
New Massive Gravity}

\author{Lucas Acito$^1$, Mariano Chernicoff$^2$, Julio Oliva$^3$, Cielo Ramirez de Arellano Torres$^3$, Matías Sempe$^4$}

\affiliation{$^1$Instituto de Física La Plata (IFLP), CONICET \& Departamento de Física Dr. Emil H. Bose, UNLP C.C. 67, (1900) La Plata, Argentina.}
\affiliation{$^2$Departamento de Física, Facultad de Ciencias, Universidad Nacional Autónoma de México, A.P. 70-542, CDMX 04510, México.}
\affiliation{$^3$Departamento de Física, Universidad de Concepción, Casilla, 160--C, Concepción, Chile}
\affiliation{$^4$ Instituto de Astronomía y Física del Espacio, (CONICET-UBA) Ciudad Universitaria, Pabellón IAFE, CABA, C1428ZAA, Argentina.}

\emailAdd{acitolucas@gmail.com} 
\emailAdd{mchernicoff@ciencias.unam.mx}
\emailAdd{julioolivazapata@gmail.com}
\emailAdd{cielo.ramirezdearellano.t@mail.pucv.cl}
\emailAdd{sempe\_100@hotmail.com}

\abstract{
We study the one-loop correction to the entropy of near-extremal black holes in three-dimensional massive gravity at the special point where the theory exhibits a unique maximally symmetric vacuum and non-constant curvature hairy black holes can achieve extremality even in the static case. Focusing on the near-horizon AdS$_2\times S^1$ geometry, we evaluate the contribution of boundary graviton modes that become exact zero-modes in the extremal limit. We show that the resulting one-loop partition function generates logarithmic corrections to the semiclassical entropy, providing a new extension to higher-curvature gravity of what has been recently obtained for near-extremal black holes in General Relativity. }

\maketitle

\section{Introduction}
It was realized several decades ago that near-extremal black holes possess an energy excess above extremality that becomes too small to support the emission of Hawking quanta with energies set by the Hawking temperature \cite{Preskill:1991tb}. Near-extremal black holes at fixed charge or angular momentum possess an energy excess above extremality that scales quadratically with the temperature, $M-M_{0}\sim T^2$, where $M_0$ is the mass of the extremal black hole. In contrast, the energy of a typical Hawking quantum scales linearly with the temperature, $\omega\sim T$, as dictated by Wien's law (see Fig. 1). Consequently, sufficiently cold black holes do not have enough energy above extremality to emit a typical Hawking quantum.

\begin{figure}[ht]\label{Fig}
\centering\begin{tikzpicture}[scale=1]
\begin{axis}[xmin=-0.11,
xmax=1.4,
ymin=-0.25,
ymax=2.8,xtick=\empty,
    ytick=\empty,
    axis lines=left,axis lines=middle
,clip=false]
\addplot[domain=0:1.3,samples=200
] {x};\addplot[blue,domain=0:1.3,samples=200] {x^2};\node[blue] at (axis cs:0.5,2.22) {\scriptsize Energy above extremality $\sim T^2$};\node[black] at (axis cs:0.64,2.4) {\scriptsize Typical energy of a Hawking quanta $\sim T$};\draw
    (axis cs:1,0)
    --
    (axis cs:1,-0.05);\node[
    below,
    font=\footnotesize
]
at (axis cs:1,0)
{$T_b$};\node[
    below,
    font=\footnotesize
]
at (axis cs:-0.035,0)
{$0$};\node[
    below,
    font=\footnotesize
]
at (axis cs:1,0)
{$T_b$};\node[
    below,
    font=\small
]
at (axis cs:1.5,0.15)
{$T$};\node[
    below,
    font=\small
]
at (axis cs:0.05,3.2)
{$M-M_0$};
\end{axis}
\end{tikzpicture}
\caption{Energy behaviour for near extremal black hole (blue line) and Hawking radiation (black line). For temperatures lower than a breakdown energy $\omega_b$, the classical black hole does not have enough energy to emit a Hawking quantum.}
\end{figure}
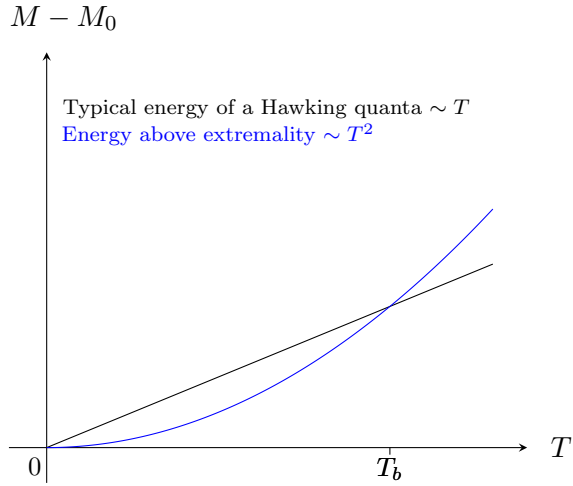

It is well-known that black hole entropy receives quantum corrections from fluctuations of gravitons and matter fields around the background geometry \cite{Sen:2012cj,Sen:2012kpz,Sen:2012dw}. Recently, these corrections have been instrumental in understanding that non-supersymmetric black holes behave as ordinary quantum systems \cite{Iliesiu:2020qvm,Ghosh:2019rcj}. In particular, it was shown in \cite{Iliesiu:2022onk} that logarithmic corrections to the entropy arise naturally from the one-loop partition function of near-extremal black holes—both for rotating solutions in vacuum General Relativity and for charged solutions in Einstein–Maxwell theory—providing a better understanding of the role played by quantum corrections regarding some thermodynamical variables. In this approach, the relevant one-loop determinant is computed by expanding the action to quadratic order around the near-horizon, near-extremal geometry, which is treated as an approximate saddle of the Euclidean path integral. In the low-temperature regime, the determinant is dominated by modes that become exact zero modes in the extremal limit. As the temperature increases, the corresponding eigenvalues are lifted by corrections linear in the temperature, giving rise to the characteristic logarithmic contributions to the entropy \cite{Iliesiu:2022onk}.

This approach relies on the assumption that the near-horizon dynamics decouples from the asymptotic region. This assumption can actually be tested and even proven in some cases. For example, for the BTZ black hole the computation performed in \cite{Kolanowski:2024zrq} correctly reproduces the low temperature behavior of the contribution to the partition function computed in \cite{Maloney:2007ud} which is one-loop exact. At low temperature the full-geometry modes that contribute to the partition function localize in the near horizon region \cite{Kolanowski:2024zrq,Acito:2025hka}, which can be also explicitly tested in the four-dimensional hyperbolic black hole \cite{Mann}. Even more, according to \cite{Kolanowski:2024zrq,Acito:2025hka,Bac:2026eqj}, this can easily be understood by studying the corresponding sector associated with Schwarzian modes. These modes arise from large diffeomorphisms that act nontrivially on the boundary of the asymptotically $AdS_2$ throat appearing in the near-horizon geometry of extremal black holes. Their dynamics is captured by an effective one-dimensional theory living at the $AdS_2$ boundary, whose action is governed by the Schwarzian derivative \cite{Maldacena:2016upp,Stanford:2017thb}. As a consequence, the low-temperature behavior of the one-loop partition function can be extracted entirely from the dynamics of these boundary degrees of freedom.

Three-dimensional gravity provides an ideal laboratory in which to test these ideas, owing to the exceptional degree of analytic control that it affords. In particular, the one-loop partition function on the BTZ black hole background can be computed exactly in the full geometry for arbitrary values of the mass and angular momentum \cite{Giombi:2008vd} (see also \cite{Witten:1988hc,Maloney:2007ud,Leston:2023ugd,Goya:2025bls}). Remarkably, the low-temperature behavior of this exact result is fully reproduced by the contribution of the Schwarzian sector associated with the near-horizon geometry, providing strong evidence for the decoupling picture described above \cite{Kolanowski:2024zrq}. Vacuum General Relativity in three dimensions does not possess local propagating degrees of freedom, and the existence of near-extremal black holes in this context necessarily relies on nonvanishing angular momentum. Consequently, the BTZ black hole furnishes a particularly clean setting in which the quantum dynamics of near-extremal horizons can be studied in the absence of bulk gravitational excitations.

In this context, another natural arena in which to address these questions is provided by three-dimensional theories of gravity with local propagating degrees of freedom. Such theories can be constructed by supplementing the Einstein-Hilbert action with suitable higher-derivative interactions while maintaining consistency at the classical and perturbative quantum levels. An important example is Topologically Massive Gravity (TMG) \cite{TMG}, in which the Einstein-Hilbert action is supplemented by a Lorentz-Chern-Simons term with a new coupling $\mu_{\mathrm{TMG}}$ of mass dimension one. Since the Lorentz-Chern-Simons term is parity odd, due to the presence of the Levi-Civita tensor $\epsilon$, TMG propagates a single massive graviton for generic values of the topological mass and cosmological constant $\Lambda$.

Another interesting example of three-dimensional massive gravity is the so-called New Massive Gravity (NMG), introduced by Bergshoeff, Hohm, and Townsend in \cite{Townsend}. This theory is parity-even and 
has the remarkable property that, despite yielding fourth-order equations, it propagates exactly two massive helicities and avoids the scalar ghost that generically plagues higher-derivative gravity. In this case the Einstein-Hilbert term is supplemented by a precise curvature square combination, with a new coupling $m$ which is the mass of the graviton. For generic values of the couplings, New Massive Gravity admits two distinct maximally symmetric vacua whose curvature radii are determined by both the bare cosmological constant $\lambda$ and the graviton mass $m$. A particularly interesting situation arises at the special point $\lambda=m^2$, where the two vacua coalesce into a unique $AdS_3$ background and the theory develops several remarkable properties. In addition to the enhancement of the linearized gauge symmetry, the space of solutions becomes larger and includes a family of non-constant-curvature black holes obeying boundary conditions that are weaker than those of Brown and Henneaux \cite{Brown:1986nw,OTT}. These geometries carry an additional integration constant $b$, usually referred to as the gravitational hair, and admit a static representative described by the blackening factor \cite{Bergshoeff:2009aq,OTT},
\begin{equation}
f(r)=\frac{r^2}{\ell^2}+br+c.
\end{equation}

Unlike the BTZ black hole, whose geometry is locally $AdS_3$, these hairy solutions owe their existence entirely to the higher-curvature dynamics of the theory. Most importantly for our purposes, they admit an extremal limit even in the absence of rotation. This feature sharply contrasts with three-dimensional General Relativity, where near-extremality necessarily requires non-vanishing angular momentum. As a consequence, the static hairy black hole provides a particularly clean setting in which to investigate the quantum thermodynamics of near-extremal horizons in a theory with propagating massive gravitons. The thermodynamic properties of these solutions are well understood. Their entropy can be computed semiclassically using Wald's formula and is precisely reproduced by Cardy's formula in the dual conformal field theory, whose central charge is twice the Brown--Henneaux  \cite{Giribet:2009qz}. In the present work, we revisit these black holes from a bulk perspective and compute the leading one-loop correction to the entropy in the near-extremal regime. Focusing on the contribution of boundary gravitons to the Euclidean path integral, we find that the semiclassical entropy $S_\text{scl}$ receives a logarithmic correction of the form
\begin{equation}
S=S_\text{scl}+\frac{3}{2}\log \left(\frac{8\pi T\ell}{G|b|}\right)+\ldots ,
\end{equation}
where $\ldots$ correspond to corrections to the semiclassical entropy of a quantum origin, which are polynomial in $T$, near-extremality. To the best of our knowledge, our manuscript represents the first work where a 1-loop correction for the entropy of near-extremal black holes has been carried out, in a higher-derivative, healthy model, with local degrees of freedom.\footnote{In \cite{SahaBaner}, a similar approach was considered in order to study the corresponding contributions due to the quadratic terms in the curvature in four dimensions in a perturbative manner, and for charged black holes in five-dimensional Einstein-Gauss-Bonnet theory see \cite{EGBus}.}

The paper is organized as follows. In Section II we review the salient features of New Massive Gravity at the special point and introduce the family of static hairy black holes that will serve as the background geometry throughout this work. We discuss their thermodynamic properties, the extremal limit, and the near-horizon geometry relevant for the low-temperature analysis. In Section III we present the general framework for computing the one-loop partition function in the near-extremal regime and identify the class of gravitational fluctuations responsible for the logarithmic corrections. Section IV contains the explicit evaluation of the temperature-induced lifting of the tensor zero-modes and the resulting contribution to the entropy. We also show that the natural candidates for vector zero-modes lead to an infinite shift in the eigenvalue when a small temperature $\mathcal{O}(T^2)$ is turned on, and therefore cannot be treated perturbatively to first order in the temperature. In Section V we provide a complementary interpretation of these modes by deriving them from a Kerr--Schild construction and establishing their relation to the Schwarzian sector of the near-horizon geometry. Additionally, from this perspective we also construct the
 rotational modes which would be valid zero-modes if they led to a finite shift in the eigenvalue for small temperature. In Section VI, we conclude with a discussion of our results and several directions for future work. Some technical details concerning the quadratic fluctuation operator of New Massive Gravity are collected in Appendix A.

\section{New Massive Gravity and Hairy Black Hole}\label{sec:HBH}

 In NMG, the Einstein-Hilbert action is supplemented by a specific combination of curvature-squared invariants. The corresponding action is given by \cite{Townsend},
\begin{align}\label{La2}
   &I_{\mathrm{NMG}}= \frac{1}{16\pi G}\int d^{3}x\sqrt{-g} \left(R -2\lambda -\frac{1}{m^2}\left(R^{\mu\nu}R_{\mu\nu}-\frac{3}{8}R^2\right)\right),
\end{align}
where $m$ is the mass of graviton and the relative coefficient $-3/8$ in the quadratic terms is essential for the nice properties of the theory, e.g. the decoupling of a ghost-like degree of freedom. The field equations coming from (\ref{La2}) read
\begin{equation}\label{eom}
G_{\mu \nu}+\lambda g_{\mu \nu}-\frac{1}{2m^2}K_{\mu \nu}=0\ ,\\
\end{equation}
where the tensor
\begin{equation}\label{elKmunu}
K_{\mu \nu}=2\nabla^2R_{\mu\nu}-\frac{1}{2}(\nabla_\mu \nabla_\nu R+g_{\mu \nu}\nabla^2 R)-8 R_{\mu \rho}R^{\rho}_{\,\,\nu}+\frac{9}{2}RR_{\mu\nu}+\frac{1}{8}g_{\mu\nu}\left(24R^{\alpha \beta}R_{\alpha \beta}-13R^2\right)\ ,
\end{equation} 
is of fourth-order in the metric, and satisfies 
\begin{equation}
K_{\mu \nu}g^{\mu\nu}=K=R_{\mu\nu}R^{\mu\nu}-\frac{3}{8}R^{2}\ ; \label{latter}
\end{equation}
i.e. its trace turns out to be proportional to the Lagrangian density from which it derives and it has only two derivatives of the metric. The latter property is crucial to prove the consistency of the 3D theory \cite{Townsend}.

 This theory propagates local massive degrees of freedom in AdS$_3$, in addition to the usual boundary excitations~\cite{Townsend}. For generic values of the coupling constants, NMG theory admits two maximally symmetric
vacua with constant curvature $R_{\ \ \lambda\rho}^{\mu\nu}=\Lambda \delta_{\lambda\rho}^{\mu\nu}$, with $\Lambda$ admitting two values

\begin{equation}
\Lambda_{\pm }=2m^{2}\left(  1\pm\sqrt{1-\frac{\lambda}{m^{2}}}\right)  \ .\label{vacua}
\end{equation}
When $m$ is large and $\lambda $ is fixed, we get
\begin{eqnarray}
\Lambda_- \simeq \lambda + \mathcal{O}(\lambda^2/m^2)\, ,\ \ \ \ \ \ \Lambda_+ \simeq 4m^2-\lambda + \mathcal{O}(\lambda^2/m^2)\, .
\end{eqnarray}
$\Lambda_+$ has to be regarded as a non-perturbative solution. On the curve of the parameter space defined by $\lambda=m^{2}$, the two vacua (\ref{vacua}) coincide and the space of solutions
is enlarged \cite{OTT}. $\lambda=m^{2}$ is called the special point of NMG. In this case, the theory around the maximally symmetric background exhibits, besides diffeomorphisms, a new gauge symmetry at linearized level; this is given by the conformal transformation (see e.g. \cite{Gabadadze:2012xv})
\begin{equation}
\delta g_{\mu\nu}(x)\,=\,\omega(x)\, \gamma_{\mu\nu}\, ,
\end{equation}
where $\gamma_{\mu\nu}$ is the metric on the constant curvature background,
and $\omega(x)$ is a local, infinitesimal parameter.

Besides the enhancement of the linearized gauge symmetry, the space of solutions becomes larger than in generic regions of the parameter space and includes a family of black hole geometries carrying an additional continuous parameter, commonly referred to as gravitational hair \cite{Bergshoeff:2009aq,OTT}. 
The static member of this family, in the Euclidean continuation is described by the metric

\begin{equation}\label{hairy}
ds^2=f(r)dt^2+\frac{dr^2}{f(r)}+r^2d\phi^2 ,
\end{equation}
where
\begin{equation}
f(r)=\frac{(r-r_+)(r-r_-)}{\ell^2} =\frac{r^2}{\ell^2}+br+c\ ,
\end{equation}
with $b=-\frac{r_++r_-}{\ell^2}$, $c=\frac{r_+r_-}{\ell^2}$ and  $2m^2=2\lambda=-\frac{1}{\ell^2}$. Notice that in asymptotically AdS spaces, it is possible to consider excitations with $m^2$ negative, provided they are above the corresponding Breitenlohner-Freedman bound \cite{BF}. The existence of the event and the Cauchy horizon requires the integration constant $b$ to be negative. We restrict the analysis to such case. 
For an appropriate range of parameters, the Lorentzian solution possesses two real horizons, an event horizon at $r=r_+$ and an inner Cauchy horizon at $r=r_-$. Unlike the BTZ black hole, whose geometry is locally AdS$_3$, the hairy black hole is not a constant-curvature spacetime and owes its existence entirely to the higher-derivative dynamics of New Massive Gravity. The horizons cover a curvature singularity at the origin since the Ricci scalar reads
\begin{equation}
  R=-\frac{6}{\ell^2}+\frac{2(r_+ +r_-)}{\ell^2r}\ .
\end{equation}
Notice that, consistently, the polynomial divergence at the origin disappears for BTZ, namely when $\ell^2 b=-(r_++r_-)=0$. The mass, temperature and entropy of the hairy black hole were computed in \cite{Giribet:2009qz}, in terms of $r_+$ and $r_-$, and they are respectively given by
\begin{equation}
M=\frac{(r_+-r_-)^2}{16G\ell^2}\ ,
\end{equation}
\begin{equation}
    T=\frac{r_{+}-r_{-}}{4\pi \ell^2}\ ,
\end{equation}
and
\begin{equation}
    S_\text{Wald}=\frac{\pi(r_+-r_-)}{2G}\ .
\end{equation}
The latter is obtained by using Wald's formula for diffeomorphism invariant, higher-curvature gravities \cite{Wald:1993nt}. This result for the black hole entropy, together with the Hawking temperature and the mass of the black hole, can be shown to satisfy the first law of black hole thermodynamics,

\begin{equation}
dM=TdS_\text{Wald}\ .
\end{equation}
Notice that there is no global charge associated to the parameter $b$ in the first law. The family of static, hairy black holes parameterized by the pair $(r_+,r_-)$, can also be parameterized by the pair of independent variables $(T,b)$. It is interesting to notice that both the entropy and mass of the extremal black hole vanish when $r_+=r_-$. For arbitrary values of $b<0$ and $T$ the outer and inner horizon radius can be written as
\begin{align}\label{rmasrmenos}
    r_+(T)&=\frac{1}{2}|b|\ell^2+2\pi \ell^2 T \equiv r_{0}+\ell\epsilon(T)  \  ,\\
    r_-(T)&= \frac{1}{2}|b|\ell^2-2\pi \ell^2 T \equiv r_{0} - \ell\epsilon(T)\ ,
\end{align}
where we have defined the extremal radius $r_0=\frac{1}{2}|b|\ell^2$ and the dimensionless, infinitesimal parameter controlling the departure from extremality  as a function of the temperature as $\epsilon(T)=\frac{1}{2\ell}(r_+-r_{-} )$, which vanishes linearly as $T\rightarrow 0$. From these, we obtain the Wald entropy and mass in terms of the temperature as
\begin{align}
  S_\text{Wald}&=\frac{2\pi^2\ell^2T}{G}\ , \\
  M&=\frac{\pi^2\ell^2T^2}{G}\ .
\end{align}
As emphasized in the Introduction, our main interest lies in the near-extremal regime. Nevertheless, it is worth mentioning that Eqs.~(17) and (18) are exact and do not rely on any low-temperature approximation. In particular, when expressed in terms of the Hawking temperature, both the entropy and mass are independent of the gravitational hair parameter $b$. The latter enters the solution only through the radius of the extremal configuration,
$r_0$, which determines the size of the $S^1$ factor in the near-horizon geometry.
%  Equations~(19) and (20) further show that departures from extremality are controlled by the dimensionless parameter $\epsilon(T)=2\pi \ell T$,
% which vanishes linearly as $T\rightarrow 0$. In the following, we shall regard $\epsilon(T)$ as a small expansion parameter and systematically organize the near-extremal geometry as a perturbation around the extremal background.

The existence of a smooth extremal limit with vanishing entropy and mass makes the static hairy black hole particularly well suited for the study of quantum corrections in the near-extremal regime. As in the case of Kerr and Reissner--Nordström black holes, the relevant low-energy dynamics is expected to be governed by the geometry of the throat that develops near the horizon as extremality is approached. Following the general strategy of Refs.~\cite{Iliesiu:2020qvm,Kolanowski:2024zrq}, we shall therefore focus on the near-horizon region of the near-extremal solution and treat the resulting geometry as an approximate saddle of the Euclidean path integral. This perspective allows one to isolate the nearly-zero modes responsible for the logarithmic corrections to the entropy and provides a controlled framework for evaluating the corresponding one-loop determinant.

\section{The one-loop functional integral and the gravitational tensor modes}\label{sec:1loop}

The partition function of a quantum field theory for a generic field $\Phi$ is obtained by the expression
\begin{equation}\label{genPI}
    Z=\int D\Phi e^{-I[\Phi]}\ ,
\end{equation}
where $I[\Phi]$ is the Euclidean action. When there is a redundancy in the description of the theory, as in (non-)perturbative quantum gravity, one can follow the Faddeev-Popov procedure in order to mod-out such redundancy, by the inclusion of a gauge-fixing term and the Faddeev-Popov (FP) ghosts. Performing and making sense of the expression \eqref{genPI} in gravity has shown to be a remarkable task. Despite such difficulties, the gravitational path-integral can be used to extract useful information about gravitational systems. Expanding to second order around a classical saddle $\Phi=\bar{\Phi}+\phi$, with $\bar{\Phi}$ the Euclidean extremum of the action and $\phi$ the quantum fluctuation, the one-loop expansion of the partition function is given by
\begin{equation}
    Z=e^{-I[\bar\Phi]}\int D\phi e^{-\int d^3x\sqrt{\bar{g}}\phi\mathcal{D}\phi}\ ,
\end{equation}
where $\mathcal{D}$ stands for a linear differential operator, acting on the Gaussian quantum field $\phi$ with an implicit tensor structure. The Gaussian integral leads to the determinant of the operator  $\mathcal{D}$ which can be written in terms of its eigenvalues $\lambda_i$, formally leading to
\begin{equation}\label{Zeigenvalues}
    Z\sim e^{-I[\bar\Phi]}\frac{1}{\sqrt{\text{det}\mathcal{D}}}\sim e^{-I[\bar\Phi]}\prod_i\frac{1}{\sqrt \lambda_i}\ .
\end{equation}
In Eq.~(\ref{Zeigenvalues}) we have used the symbol $\sim$ rather than an equality sign because the Gaussian functional integral produces an overall factor that depends on the choice of measure. Since this factor is independent of the temperature, it plays no role in the extraction of the low-temperature one-loop corrections and will be ignored in what follows. The free energy is determined by the logarithm of the partition function, which yields

\begin{equation}
\log Z=-I[\bar{\Phi}]
-\frac{1}{2}\sum_i \log \lambda_i
+\ldots ,
\end{equation}
where the ellipsis denotes quantum contributions that are subleading at low-temperature expansion. Our goal is to evaluate the one-loop correction to the partition function around the near-extremal limit of the saddle \eqref{hairy}. To this end, we regard the near-extremal geometry as a perturbation of the extremal background and expand both the classical solution and the fluctuation operator in powers of the temperature. The eigenvalues of $\mathcal{D}$ can then be written as
\begin{equation}
\lambda_i=\bar{\lambda}_i+\delta\lambda_i,
\qquad
\delta\lambda_i=\mathcal{O}(T),
\end{equation}
where $\bar{\lambda}_i$ are the eigenvalues of operator $\mathcal{D}$ on the extremal geometry. Accordingly,

\begin{equation}
\log Z=
-I[\bar g_{\mu\nu}+T,\delta g_{\mu\nu}]
-\frac{1}{2}
\sum_i
\log\left(\bar{\lambda}_i+\delta\lambda_i\right).
\end{equation}
For modes with non-vanishing extremal eigenvalues, $\bar{\lambda}_i\neq 0$, one finds
\begin{equation}
\log\left(\bar{\lambda}_i+\delta\lambda_i\right)=\log \bar{\lambda}_i
+
\frac{\delta\lambda_i}{\bar{\lambda}_i}
-\frac{1}{2}
\left(
\frac{\delta\lambda_i}{\bar{\lambda}_i}
\right)^2
+\mathcal{O}(T^3).
\end{equation}
Since $\delta\lambda_i\propto T$, the contribution of these modes generates only analytic, power-law corrections in the temperature. Consequently, they are subleading in the low-temperature regime and do not contribute to the logarithmic terms that are the focus of this work.

The situation is qualitatively different for modes that become exact zero modes in the extremal limit. For such modes, $\bar{\lambda}_i=0$, and the previous expansion breaks down. Instead, one obtains
\begin{equation}
\log(\delta\lambda_i)
\sim
\log T,
\end{equation}
which dominates over any power-law correction as $T\to0$. Retaining only these contributions, the partition function reduces to

\begin{equation}
\log Z=
-I[\bar g_{\mu\nu}+T,\delta g_{\mu\nu}]
-\frac{1}{2}
\sum_i^{(0)}
\log(\delta\lambda_i)
+\ldots ,
\end{equation}
where the sum runs over the zero modes of $\mathcal{D}$ on the extremal $AdS_2\times S^1$ background, and the ellipsis denotes terms that are subleading in the near-extremal expansion.

The near-extremal geometry lifts both the zero eigenvalues and their degeneracy. To first order in the temperature, the corresponding shift is given by
\begin{equation}
\delta\lambda_i=\int d^3x\sqrt{\bar g}
h^i_{\mu\nu}
\delta\mathcal{D}\left(h^i
\right)^{\mu\nu},
\label{deltalambda}
\end{equation}
where $\delta\mathcal{D}$ denotes the part of the quadratic fluctuation operator that is linear in the temperature, and $h^i_{\mu\nu}$ is the $i$-th normalized zero mode of the extremal background. The explicit form of the NMG fluctuation operator is presented in Appendix~A.

Now, we implement these ideas in a precise manner, in the context of the near-extremal, near-horizon geometries of the static hairy black hole in NMG. As it is done in General Relativity, we will focus on the near-horizon geometry of the near-extremal Euclidean black hole, and consider such configuration as an approximate saddle, disregarding terms $\mathcal{O}(T^2)$. The change of coordinates
\begin{equation}
    r=r_{+}+\ell\epsilon(T)(\cosh{\eta}-1) , \qquad t=-i\frac{\ell}{\epsilon(T)}\tau,
\end{equation}
followed by the expansion $T\rightarrow 0$, allows zooming into the near-horizon region leading to the metric
\begin{equation}\label{appsaddle}
    g_{\mu\nu}=\bar{g}_{\mu\nu}+T\delta g_{\mu\nu}+\mathcal{O}(T^2) \ .
\end{equation}
Notice that we have made explicit the linear dependence in $T$, that is,

\begin{equation}
    d\bar{s}^2=\ell^2(\sinh{\eta}^2d\tau^2+d\eta^2)+r_{0}^2d\phi^2
\end{equation}
and 
\begin{equation}
\delta g_{\mu\nu}dx^\mu dx^\nu= 2\pi r_{0} \ell^2 \cosh{\eta} \  d\phi^2\ .
\end{equation}

The metric $\bar{g}_{\mu\nu}$, is the near-horizon geometry of the extremal black hole, and it corresponds to an $AdS_2\times S^1$ geometry without any non-trivial fibration since our 2+1-dimensional, near-extremal black hole, is static. The radius of the $S^1$ factor is fixed by the gravitational hair parameter $b$.

Our goal is to determine the one-loop contribution of gravitational fluctuations to the near-extremal black hole entropy. In principle, the functional integral of New Massive Gravity receives contributions from both the local massive graviton and the boundary graviton sectors. However, in the low-temperature regime $T\ll m$, the propagating bulk modes are expected to be suppressed by the graviton mass. Consequently, the dominant contribution to the partition function should arise from the gapless boundary degrees of freedom associated with large diffeomorphisms. Therefore, considering also the universality arguments of \cite{Karan:2022dfy,Maulik:2024dwq} and the analysis of near-extremal black holes in Einstein gravity, we shall focus on the sector generated by metric fluctuations of the form,
\begin{equation}
h_{\mu\nu}=\mathcal{L}_{\xi}g_{\mu\nu}\ ,
\end{equation}
for which the perturbation $h_{\mu\nu}$ is normalizable, while the vector field $\xi^\mu$ is not. These modes correspond to physical boundary gravitons and constitute the gravitational realization of the Schwarzian degrees of freedom that govern the low-energy dynamics of the $AdS_2$ throat. In the BTZ black hole, this sector is known to reproduce the exact low-temperature behavior of the one-loop partition function, despite being entirely determined by the near-horizon geometry. One of the main goals of the present work is to investigate whether this mechanism continues to hold beyond GR.

To isolate these modes, we impose the de Donder gauge condition
\begin{equation}\label{dD}
\nabla^\mu h_{\mu\nu}-\frac{1}{2}\nabla_\nu h^\mu_{\ \mu}=0,
\end{equation}
and further restrict our attention to traceless fluctuations,
\begin{equation}\label{traza}
h^\mu_{\ \mu}=0.
\end{equation}
Let us notice that beyond GR, at the classical level, $h^\mu_{\ \mu}$ could be a propagating degree of freedom. At the quantum level, even in GR, considering the function integral over both the traceless part of $h_{\mu\nu}$ and its trace $h^\mu_{\ \mu}$ leads to the problem of the conformal mode \cite{Gibbons:1978ac}.

The conditions \eqref{dD} and \eqref{traza} are automatically satisfied by the Schwarzian zero modes considered below. The resulting perturbations admit a smooth extension from the AdS$_2$ throat to the full black hole geometry and are therefore expected to capture the leading logarithmic corrections to the entropy.
According to their transformation properties on the AdS$_2$ factor of the near-horizon geometry, the fluctuations can be classified into tensor, vector, and scalar modes. The tensor sector contains the Schwarzian zero modes whose eigenvalues are lifted linearly away from extremality and therefore give rise to the logarithmic corrections discussed in the Introduction. We shall show that these modes provide the leading contribution to the one-loop partition function. The vector, or rotational, sector will be analyzed separately and shown to yield subleading corrections in the near-extremal expansion.

\section{Results for the tensor and would-be vector modes}

\subsection{Tensor modes}
The tensor modes corresponding to the degrees of freedom described by large diffeomorphism are given by \cite{Campo,Sen:2012kpz}
\begin{equation}
  h^{(n)}_{\mu\nu}dx^\mu dx^\nu= \left(\frac{2(n^2-1)|n|}{|b|}\right)^{1/2}\ e^{i n\tau}\tanh^{|n|}\left(\frac{\eta}{2}\right) \left(i \frac{n}{|n|}d\tau+\frac{d\eta}{\sinh{\eta}} \right)^2 .
  \label{tensor_mode}
\end{equation}
These modes satisfy the correct conjugation property since $\left(h^{(n)}_{\mu\nu}\right)^*=h^{(-n)}_{\mu\nu}$ and for $|n|\geq 2$ they are normalizable. These modes allow to reconstruct a real field $h_{\mu\nu}$ as a superposition of the $\left(h^{(n)}_{\mu\nu}\right)$, which are orthonormal.
As explained before, they are generated by large diffeomorphisms $\xi^{(n)}$ acting on the $AdS_{2}$ submanifold, that is $h^{(n)}_{\mu\nu}=\mathcal{L}_{\xi^{(n)}}\bar g_{\mu\nu}$ where\footnote{Notice that diffeomorphisms which asymptotically go to the Brown-Henneaux vector field, globally defined in AdS$_3$ and the exterior of the Lorentzian static BTZ, were explicitly constructed in a different context in \cite{Anabalon:2010um}.}
% \begin{equation}
%   \xi^{\mu}\partial_{\mu} \sim e^{i n\tau}\tanh^{|n|}{\left(\frac{\eta}{2}\right)}\left(i\frac{\left(-1+n^2+|n|\cosh{\eta}+\cosh{\eta}^2\right)}{ n\sinh^2{\eta}}\partial_{\tau} + \frac{|n|+\cosh{\eta}}{\sinh{\eta}}\partial_{\eta}\right)\ .
% \end{equation}
\begin{equation}
  \xi^{\mu}\partial_{\mu}=\mathcal{N}_n e^{i n \tau}\frac{\tanh^{|n|}{\left(\frac{\eta}{2}\right)}}{\sinh{\eta}}\left(\frac{i(\sinh^2{\eta}+|n|(|n|+\cosh{\eta}))}{n\sinh{\eta}}\partial_{\tau}+(|n|+\cosh{\eta})\partial_{\eta}\right),
\end{equation}
with $\mathcal{N}_{n}$ given by the following normalization factor
\begin{equation}
  \mathcal{N}_{n}=\frac{ |n| }{\sqrt{2|b|(n^2-1)}\ell^2}.
\end{equation}
These tensor modes are transverse and traceless on the background $T=0$ geometry. The integration required to compute $\delta\lambda^{(n)}$ in \eqref{deltalambda} for the tensor modes can be performed in an explicit manner for arbitrary $n$, and leads to
\begin{equation}
  \delta\lambda^{(n)}=\frac{4|n|\pi}{|b|\ell^2}T\ .
  \label{deltalambdasol}
\end{equation}
As discussed above, we have parametrized the near-extremal black holes as a function of the temperature and the gravitational hair $b$. The latter determines the radius of the extremal horizon, and explicitly appears in the correction to the zero-modes.  
\subsection{The would-be vector modes}
Transverse and traceless vector modes (also called rotational modes of the theory) can be constructed using the Killing direction of the background $\partial_\phi$, as
\begin{equation}
  h^{(n)}_{\mu\nu}dx^{\mu}dx^{\nu}=\frac{\sqrt{|n||b|}\ell}{2} e^{i n\tau}\tanh^{|n|}\left(\frac{\eta}{2}\right) \left(i \frac{n}{|n|} d\tau+\frac{d\eta}{\sinh{\eta}} \right)\otimes_s d\phi\ , 
  \label{vector_mode}
\end{equation}
where $\otimes_s$ denotes the symmetrized tensor product. These modes are annihilated by the generalized Lichnerowicz operator on the background geometry. Even more, they are orthonormal on such geometry for $n\neq 0$. These modes are obtained by large diffeomorphisms acting on the $S^{1}$ submanifold that take the form
\begin{equation}
  \xi^{\mu}\partial_{\mu} \sim e^{i n\tau}\frac{\tanh^{|n|}\left(\frac{\eta}{2}\right)}{|n|}\ \partial_{\phi}\ ,
\end{equation}
and the vector field itself is not square integrable on the near-horizon extremal geometry, which is why we associate the modes $h_{\mu\nu}$ to large diffeomorhisms. These results would suggest to interpret the vector modes as important for the computation of the one-loop partition function. Nevertheless, for these modes, $\delta\lambda_i$ diverges after integrating the density $\sqrt{\bar g}h^i_{\mu\nu}\delta\mathcal{D}\left(h^i\right)^{\mu\nu}$, which before integration goes as $\mathcal{O}(T^2)$. Therefore, in this setup, namely for the near-extremal, static hairy black hole in NMG, the contribution of the vector modes has to be disregarded.

\subsection{One-loop corrected entropy}
Using the result we obtained from the one loop contribution in \eqref{deltalambdasol}, we can now calculate the relevant sector of the partition function that contribute at low temperature given by
\begin{equation}
  Z
=
\exp\left[
\frac{4\pi^2 \ell^2}{\beta}
-\sum_{n=2}^{\infty}
\log\left(
\frac{4\pi n}{\beta b \ell^2}
\right)
\right].
\end{equation}
Using zeta function regularization we can re-obtain the famous $\frac32$ contribution that we expect
\begin{equation}
  Z
=
\exp\left[
\frac{4\pi^2 \ell^2}{\beta}
-\frac12\log\left(2 \pi \right)+
\frac32\log\left(
\frac{4\pi }{\beta b \ell^2}
\right)
\right].
\end{equation}
we can obtain the entropy from here using that 
\begin{equation}
  S=(1-\beta \partial_\beta) \log Z
\label{Sdef}
\end{equation}
utilizing this formula in our case, we obtain that the grand canonical entropy is given by
\begin{equation}
S =
\frac{1}{2}
\left[
3 + \frac{16\pi^2}{\beta}
- \log(2\pi)
+ 3\log\left(
\frac{4\pi}{b\ell^2\beta}
\right)
\right].
\end{equation}
this gives a competition between the classical linear contribution and the quantum logarithmic contribution that dominates at low enough temperature. Nevertheless we can take the inverse Laplace transform from the $ Z$ to obtain the energy distribution given by
\begin{equation}
\rho(E)
=
\frac{4}{\sqrt{\pi}}
\left(\frac{1}{b\ell^2}\right)^{3/2}
\sinh\left(4\pi\sqrt{E}\right).
\end{equation}
which goes to zero at $E\to0$ as expected from a non-supersymmetric black hole energy distribution.

Indeed, upon restoring Newton's constant, the one-loop corrected entropy takes the form
\begin{equation}
S=S_\text{scl}+\frac{3}{2}\log \left(\frac{4\pi T\ell}{G|b|}\right)+\ldots,
\end{equation}
from which we can identify the quantum temperature scale
\begin{equation}
T_q=\frac{G|b|}{4\pi\ell},
\end{equation}
which indeed has mass dimension one, since $b$ has mass dimension one as well. This result follows directly from a careful dimensional analysis in which the metric components are dimensionless, while the coordinates carry dimensions of length. With these expressions at hand, one concludes that the regime of validity of our computation is
\begin{equation}\label{regime}
   \frac{3G}{4\pi^2 \ell^2}\, W\!\left(\frac{\pi \ell |b|}{3}\right)<T<T_q
\end{equation}
where $W$ stands for the Lambert function. Here, the upper bound ensures that we are in the perturbative regime, since $\delta\lambda\sim\frac{T}{T_q}$ remains smaller than 1, while the lower bound ensures that the one-loop corrected entropy remains positive since below this bound the one-loop quantum correction competes with the semiclassical result, and the bound exists regardless of the fact that the semiclassical entropy of the extremal black hole vanishes. The family of hairy black holes we are considering in this work exist for arbitrary values of $b$, which as explained above, must be negative in order to ensure the existence of an extremal case. Namely, $|b|$ can take arbitrarily large values. Since for large $|b|$ the $W$-Lambert function diverges logarithmically, it is always possible to achieve the regime of validity imposed by \eqref{regime}. Even though the hair parameter $|b|$ is not related to a global charge of the black hole, it has a clear geometric meaning since it sets the size of the extremal black hole via $r_0=\ell^2|b|/2$.

\section{Zero-modes of extremal black holes in NMG from a Kerr-Schild construction}

The gravitational tensor modes obtained in the previous Section can also be derived from the Kerr–Schild metric around the hairy black hole. The computation of these modes was first carried out in the context of the BTZ black hole in \cite{Acito:2025hka}, where the relation holds throughout the entire geometry, whereas for generic metrics it seems to be fulfilled only in the near horizon extremal limit. The underlying idea of this ansatz is to describe perturbations that propagate along some privileged null directions, thereby preserving the causal structure of the background (for a complete review, see \cite{Stephani:2003}).

We begin by computing the null geodesics of a generic Euclidean axisymmetric metric of the form
\begin{equation}
    ds^2=f(r) dt^2+\frac{1}{f(r)}dr^2+r^2 d\phi^2\ .
    \label{metricHBH}
\end{equation}
This spacetime admits two Killing vectors, $\partial_t$ and $\partial_\phi$, leading to the conserved quantities
\begin{align}\label{consE}
    E&=g_{\mu \nu}\,\delta^\mu_{\ t}\,\dot{x}^\nu=f(r)\,\dot{t} \ ,\\
    \label{consL}L&=g_{\mu \nu}\,\delta^\mu_{\ \phi}\,\dot{x}^\nu=r^2 \dot{\phi}\ ,
\end{align}
along the geodesic motion. We are looking for null geodesics $\dot{x}^\mu$, where the dot indicates differentiation with respect to an affine parameter. Hence, they satisfy the condition
\begin{equation}
    g_{\mu\nu}\dot{x}^\mu\dot{x}^\nu=f(r) \dot{t}^2+\frac{1}{f(r)} \dot{r}^2+r^2 \dot{\phi}^2=0\ .
\end{equation}
Using the conserved quantities \eqref{consE} and \eqref{consL}, this condition becomes
\begin{equation}
    0=\frac{E^2}{f(r)}+\frac{1}{f(r)} \dot{r}^2+\frac{L^2}{r^2}\ ,
\end{equation}
and therefore, the null geodesics can be written in the following form
\begin{equation}
    \dot{x}^\mu \partial_\mu = \frac{E}{f(r)}\,\partial_t \pm i\sqrt{E^2+L^2 \frac{f(r)}{r^2}}\,\partial_r +\frac{L}{r^2}\,\partial_\phi\ .
    \label{nullgeo}
\end{equation}
In this Euclidean geometry, the $\pm$ signs lack the standard Lorentzian interpretation of in-going or out-going directions across the horizon; consequently, both branches must be considered.

Among these null geodesics, there exists a distinguished family corresponding to the Principal Null Directions (PNDs). These directions are defined along the principal axes of the Ricci tensor. Instead of diagonalizing the Ricci tensor directly, it is convenient to introduce the traceless Ricci tensor,
\begin{equation}
  S_{\mu\nu}= R_{\mu\nu}-\frac{1}{3}\,g_{\mu\nu}\,R\ ,
\end{equation}
and diagonalize it, since its eigenvectors are identical to those of the Ricci tensor by definition, i.e., 
\begin{equation}
  S^\mu_{\ \nu}k^{\nu}=\lambda \,k^\mu\ .
\end{equation}
The eigenvectors $k^\mu$ thus indicate the principal directions. For the metric \eqref{metricHBH}, the traceless Ricci tensor reads
\begin{equation}
    S_{\mu \nu} dx^\mu dx^\nu=\frac{f'(r)-r f''(r)}{6 r f(r)} (dr^2-2 f(r) r^2 d\phi^2+f(r)^2 dt^2)\ .
\end{equation}
A direct computation shows that the principal directions require the azimuthal component to vanish, $k^\phi=0$, on the static geometry. Thus, we can construct the principal null geodesics by setting $L=0$ in \eqref{nullgeo}. The associated vector fields are
\begin{equation}
  k^\mu\partial_\mu = \frac{E}{f(r)}\,\partial_t \pm i|E|\partial_r \ ,
  \label{PNG0}
\end{equation}
where for both signs the corresponding eigenvalue is
\begin{equation}
  \lambda=\frac{f'(r)-r f''(r)}{6 r}\ .
\end{equation}

\textbf{Kerr-Schild construction.} Having obtained these two null geodesics, we can construct a Kerr-Schild ansatz around the background metric\footnote{The Kerr-Schild ansatz around generic, static, circularly symmetric solutions that accommodate asymptotically AdS and Lifshitz black holes in NMG has been explored in \cite{Ayon-Beato:2014wla}.} \eqref{metricHBH}. Here, we define the Kerr-Schild metric on a curved base space, keeping the notation $g_{\mu\nu}$ for the base metric \eqref{metricHBH},
\begin{equation}
    \tilde{g}^{\mu\nu}= g^{\mu\nu} + \chi\: k^\mu k^\nu\ ,
    \label{Kerr_Schild_metric}
\end{equation}
where $\chi$ is a scalar field and $k^\mu$ is simultaneously a null and geodesic vector with respect to both the hairy black hole background and the full Kerr-Schild metric
\begin{equation}\label{KS}
    \tilde{g}_{\mu\nu} k^\mu k^\nu=g_{\mu\nu}k^\mu k^\nu=0 \qquad \text{and}\qquad k^\mu\tilde{\nabla}_\mu k^\nu=k^\mu\nabla_\mu k^\nu=0 \ ,
\end{equation}
with $\tilde{\nabla}_\mu$ denoting the covariant derivative compatible with $\tilde{g}_{\mu\nu}$. The null geodesic vectors obtained above naturally satisfy (\ref{KS}).
Defining $k_\mu\equiv g_{\mu\nu}k^\nu$, the inverse metric takes the form
\begin{equation}
    \tilde{g}_{\mu\nu}= g_{\mu\nu} - \chi\: k_\mu k_\nu\ .
\end{equation}
Having determined $k_\mu$, it remains to solve for the scalar field $\chi$ in the Kerr-Schild metric \eqref{Kerr_Schild_metric}. This can be achieved by imposing the transverse gauge condition,
\begin{equation}
    \nabla^\mu \left(\chi k_\mu k_\nu\right)=0 \quad\Longrightarrow\quad k_\mu \nabla^\mu \chi \pm i\,|E|\frac{\chi}{r}=0 \ ,
    \label{k_transverse}
\end{equation}
where we have used the fact that $\nabla^\mu k_\mu=\pm i\,|E|/r$. Consequently, the equation for $\chi$ reduces to a simple first-order differential equation. We will now focus on a metric with $g_{tt}=g_{rr}^{-1}=f(r)$ of the form
\begin{equation}
  f(r)=\frac{(r-r_-)(r-r_+)}{\ell^2}\ .
\end{equation}
Using the ansatz $\chi(t,r)=e^{i E\,t}\,R(r)$ for the scalar function, one immediately finds
\begin{equation}
  R(r)=\mathcal{N}\frac{(r-r_-)^{\pm \frac{\ell^2|E|}{r_+-r_-}} (r-r_+)^{\pm \frac{\ell^2|E|}{r_--r_+}}}{r}\ ,
\end{equation}
where $\mathcal{N}$ is a normalization constant. Notice that although we are considering both principal null geodesics obtained in \eqref{PNG0}, there is only one regular solution at the outer horizon $r=r_+$, which corresponds to the minus sign in \eqref{PNG0}. Identifying the energy with the Matsubara frequencies, $E=2\pi n T$, we obtain the Kerr-Schild ansatz
\begin{equation}
  h_{\mu\nu}^{(n)}\,dx^\mu dx^\nu=\mathcal{N}(r-r_-)^{-\frac{2\pi\,T\,|n|\,\ell^2}{r_+-r_-}} (r-r_+)^{\frac{2\pi\,T\,|n|\,\ell^2}{r_+-r_-}}\frac{2\pi\,T}{r}\left(\frac{n}{|n|}\,dt - \frac{i}{f(r)}\,dr \right)^2\ ,
  \label{KS_schwarzian}
\end{equation}
where we identify the Kerr-Schild ansatz as the tensor perturbation $h_{\mu\nu}^{(n)}=\chi k_{\mu} k_{\nu}$.

Lastly, we can define the only additional vector orthogonal to both principal null directions using the azimuthal Killing vector,
\begin{equation}
  P^\mu\partial_\mu=\partial_\phi \ .
\end{equation}
Hence, following the Newman-Penrose construction, another possible ansatz can be built from a combination of the spacetime triads. Since we established that only one of the principal geodesics yields a well-defined mode, we propose the ansatz
\begin{equation}
  \tilde{h}_{\mu\nu}=\psi_0\,k_{\mu}k_{\nu}+\psi_1\,k_{(\mu}P_{\nu)}\ ,
\end{equation}
where $\psi_0$ and $\psi_1$ are scalar functions. We identify this ansatz as the rotational mode, as it is the only one with a component in the $\phi$-direction. Imposing the transverse gauge condition and focusing on axisymmetric scalar functions independent of $\phi$, we obtain
\begin{equation}
  \left(k_\mu\nabla^\mu\psi_0+\psi_0\nabla^\mu k_\mu\right)k_\nu + \frac12\left(k_\mu\nabla^\mu\psi_1+\psi_1\nabla^\mu k_\mu \right)P_\nu=0\ .
\end{equation}
Since $k_\nu$ and $P_\nu$ have support in different components, each of these terms must vanish independently. Thus, we immediately find that $\psi_0=\psi_1=\chi$. Thus, since the first part of this mode is already contained in the previous mode \eqref{KS_schwarzian}, we identified the purely rotational mode as
\begin{equation}
  \tilde{h}_{\mu\nu}=\chi\,k_{(\mu}P_{\nu)}\ .
  \label{KS_rot}
\end{equation} 

\textbf{Near-horizon extremal limit.} We now show that the modes \eqref{KS_schwarzian} precisely correspond to the zero modes obtained in the near-horizon extremal (NHE) limit of the hairy black hole. As explained in Section.~\ref{sec:HBH}, in the near-extremal regime we expand the horizon radii in powers of the temperature $T$,
\begin{equation}
    r_+=r_0+2\pi \ell^2 T +\mathcal{O}(T^2)\:,\qquad r_-=r_0-2\pi \ell^2 T +\mathcal{O}(T^2)\ .
\end{equation}
Applying the coordinate transformations
\begin{equation}
        r=r_+ +T\:2\pi\:\ell^2 (\cosh\eta-1) \:,\qquad
        t=\frac{1}{2 \pi T}\:\tau \ , 
\end{equation}
and keeping the azimuthal coordinate $\phi$ unchanged since the black hole is non-rotating, we find that the Kerr-Schild ansatz \eqref{KS_schwarzian} in the NHE limit takes the form
\begin{equation}
    h^{(n)}_{\mu \nu}dx^\mu dx^\nu= -n^2\frac{\mathcal{N}_n}{r_0}\, e^{i n \tau } \tanh\left(\frac{\eta}{2}\right)^{|n|} \left(i\frac{n}{|n|}\,d\tau + \frac{d\eta}{\sinh\eta}\right)^2\ .
    \label{schwarian_NHNE}
\end{equation}
which coincides with the zero modes found in the near-horizon analysis \eqref{tensor_mode}. On the other hand, computing the NHE limit of the rotational mode \eqref{KS_rot} yields
\begin{equation}
  \tilde{h}_{\mu\nu}dx^\mu dx^\nu=-i\,\mathcal{N}_n\, r_0 \, e^{i n \tau} \,  \tanh^{|n|}\left(\frac{\eta}{2}\right)\,\left(i\frac{n}{|n|} d\tau + \frac{d\eta}{\sinh\eta}\right)\otimes_s d\phi \ .
\end{equation}
This mode also coincides with the previously obtained vector mode~\eqref{vector_mode}. As we have already shown, these modes are subleading and therefore do not contribute to the logarithmic correction to the entropy.

In conclusion, we find that the one-loop gravitational perturbations studied in Sec.~\ref{sec:1loop} emerge here from a Kerr-Schild ansatz, which is a purely geometric construction. The fact that the Kerr-Schild ansatz could serve as a generator of physical Schwarzian modes was first noticed for BTZ black holes in \cite{Acito:2025hka}, where one can extract the tensor mode in the full geometry from this equivalence. In the present case, however, this equivalence between frameworks can only be verified in the NHE limit. Although the Kerr-Schild construction \eqref{KS_schwarzian} is performed in the full NMG geometry, these modes (which should be eigenfunctions of the Lichnerowicz operator) only become exact eigenfunctions in the NHE limit. Nevertheless, a modified ansatz for \eqref{KS_schwarzian} (for instance, taking into account that the NMG metric is not globally conformally invariant) could yield the correct eigenfunctions for the full geometry; however, such contributions would be subleading and fall beyond the scope of this work.

Lastly, the results of this section can be interpreted within the framework of the \textit{double copy}, which establishes a correspondence between perturbative gravity and gauge theories \cite{Monteiro:2014}. For BTZ black holes, this framework allows the BTZ solution to be interpreted as a gauge field configuration in an $\text{AdS}_3$ background \cite{Carrillo:2018}. While this duality has not been extensively explored in NMG, our findings point toward a double copy construction involving an Abelian gauge field defined on the background geometry. Specifically, our results suggest that a vector perturbation of this gauge field, given by $A_\mu=\chi\,k_\mu$, reproduces the Schwarzian modes of the hairy black hole, at least in the NHE limit.

\section{Final Remarks}
We have computed the one-loop correction to the entropy of near-extremal static black holes in New Massive Gravity (NMG), considering as background configurations, solutions that exist at the special point where the theory admits a unique maximally symmetric vacuum. Our analysis focused on the sector of fluctuations shared by both General Relativity and this well behaved higher-derivative theory. We have shown that the leading correction to the semiclassical low-temperature entropy takes the form
$\frac{3}{2}\log\left(\frac{T}{b}\right)+\cdots$, where $b$ denotes the gravitational hair parameter introduced in \cite{OTT} (see also \cite{Chernicoff:2020kmf,Donnay:2020yxw}). As in GR, these modes become zero-modes of the generalized Lichnerowicz operator in the extremal limit, and their eigenvalues are lifted linearly with the temperature $T$ in the tensor sector. We have also argued that, for the vector modes constructed in the near-horizon, near-extremal geometry, no linear correction in $T$ arises in the eigenvalues of the corresponding extremal zero modes, and even more, we have identified the origin for both modes from a Kerr-Schild construction. In addition, there may exist further zero modes associated with the fourth-order operator $\mathcal{D}$. Since such modes would be intrinsic to the massive graviton sector, we expect them to remain gapped and therefore to give subleading contributions in the low-temperature regime. A definitive verification of such result lies beyond the scope of the present work. In this context, it is worth mentioning that even though the background geometry is $AdS_2\times S^1$, it still is a difficult task to decouple the equations for the eigenvalue problem of the generalized Lichnerowicz operator which is of fourth order. Notice also that we have not explicitly considered the contribution to the partition function of the FP ghosts. Since their contribution vanishes in GR, and here in the context of NMG we are fixing exactly the same diffeomorphism invariance in the same manner as in GR, it is natural to expect that their contribution will vanish here as well.

Several interesting questions remain open. The static black hole studied here arises as the non-rotating limit of a more general three-parameter family of hairy rotating black holes characterized by the mass, angular momentum, and gravitational hair, $(M,J,b)$. Within this family, (near-)extremality can be reached in two distinct ways: either by tuning the hair parameter to a critical value $b=b_{\text{ext}}$ for fixed $(M,J)$, or by tuning the angular momentum to $J=J_{\text{ext}}$ for fixed $(M,b)$ (see \cite{Donnay:2020yxw,Giribet:2010ed}). It would be interesting to investigate the one-loop entropy corrections in both near-extremal regimes. In such cases, one naturally expects the appearance of rotational vector-modes of the metric perturbation $h_{\mu\nu}$, supported by the background angular momentum. It would also be worthwhile to determine whether these families of modes can be extended to the full black hole geometry in a normalizable manner. Furthermore, in \cite{Chernicoff:2024lkj} we identified an infinite class of higher-curvature theories, at arbitrary order in the curvature, that admit the same hairy black-hole solution. These theories provide natural settings in which analogous one-loop corrections could be studied. They remain sufficiently close to the NMG case in that they describe black holes with a microscopic entropy interpretation and possess an enhanced conformal symmetry around the unique maximally symmetric vacuum. Since the quasinormal modes of conformally coupled scalar fields on the hairy black hole can be obtained in a closed manner \cite{Chernicoff:2020kmf}, it would be interesting to explore whether the Denef, Hartnoll and Sachdev (DHS) approach, which computes the one-loop partition functions by expressing functional determinants as products over the quasinormal mode spectrum of the Lorentzian black-hole background \cite{DHS}, can be extended to the present NMG setup (see also \cite{Mukherjee:2024nhx}).

Finally, a complete understanding of the one-loop partition function may benefit from reformulating NMG in terms of an auxiliary tensor field $f_{\mu\nu}$, which reduces on shell to the Schouten tensor but acquires independent fluctuations in the quantum theory. Preliminary results indicate that, upon expanding the action to quadratic order in the fluctuations $(\delta g_{\mu\nu},\delta f_{\mu\nu})$, nontrivial mixing terms between the metric and auxiliary field perturbations appear. Consequently, the generalized Lichnerowicz operator must first be diagonalized before the one-loop partition function can be addressed in this formulation. This framework is also the natural setting for studying the effective deformed Schwarzian theory, since the boundary terms required for a well-posed variational principle and for the finiteness of the action on asymptotically AdS spacetimes are already known \cite{HohmTonni}. We hope to report on these developments in future work.

\acknowledgments
We thank Alejandro Alvarado, Andrés Anabalón, Gastón Giribet, Nicolás Grandi, Marcelo Oyarzo, Gabriel Ortega and Jorge Urbina for useful discussions, comments on the manuscript and work along similar lines. The work of M.C. is partially funded by DGAPA-UNAM grant IG101326. This work is supported in part by the FONDECYT grants 1230853, 1242043, 1250133, 1262452 and  126241.

\appendix

\section{Second-order expansion of the NMG action}
Let us examine the second-order expansion of the New Massive Gravity action (\ref{La2}) around a classical solution\footnote{Note that the classical solution $g^{0}_{\mu\nu}$ should not be confused with the extremal, classical solution $\bar{g}_{\mu\nu}$. Here we are expanding around a solution that solves the field equations at arbitrary temperature $T$, not just at extremality.} $g_{\mu\nu}=g^{0}_{\mu\nu}+h_{\mu\nu}$
\begin{equation}
  \left. I[g]\right|_{g=g^{0}+h}
  = I[g^{0}]
  + \int d^3x\left.\frac{\delta I}{\delta g^{\mu\nu}(x)}\right|_{g=g^{0}}h^{\mu\nu}(x)+\frac{1}{2}\int d^3x  d^3y \ h^{\lambda \rho}(x)\left.\frac{\delta^2 I}{\delta g^{\lambda \rho}(x)\delta g^{\mu\nu}(y)}\right|_{g=g^{0}}h^{\mu\nu}(y)+\ldots
\end{equation}
Since $g^{0}_{\mu\nu}$ is a solution to the field equations, the first order contribution vanishes on-shell. Thus, we see that the action can be schematically rewritten as
\begin{equation}
  \left.I[g]\right|_{g=g^{0}+h}=I[g^{0}]+h^{\lambda\rho}(\mathcal{D}[h])_{\lambda\rho}, \qquad \text{where} \ \ (\mathcal{D}[h])_{\lambda\rho}=\left.\frac{\delta^2 I}{\delta g^{\lambda \rho}\delta g^{\mu\nu}}\right|_{g=g^{0}}h^{\mu\nu}.
\end{equation}
We can decompose this operator acting on the perturbation by the number of derivatives that appear, namely
\begin{equation}
    (\mathcal{D}[h])_{\lambda\rho}=(\mathcal{D}^{(0)}[h])_{\lambda\rho}+(\mathcal{D}^{(2)}[h])_{\lambda\rho}+(\mathcal{D}^{(4)}[h])_{\lambda\rho}.
\end{equation}
Each one of them is self-adjoint, for appropriate boundary conditions. Taking the gauge condition (\ref{dD}), we find these operators to be
\begin{align}
    (\mathcal{D}^{(0)}[h])_{\mu\nu} =& -\frac{\lambda}{2} h_{\mu\nu}\\
    (\mathcal{D}^{(2)}[h])_{\mu\nu} =& -\sigma R_{\lambda\left(\mu\right.}h_{\left. \nu \right)}^{\ \ \lambda}+\frac{\sigma}{4}\left[\left(R h_{\mu\nu}+\Box h_{\mu\nu}\right)+2R_{\mu\lambda\nu\rho}h^{\lambda\rho}-\frac{g_{\mu\nu}}{2}\left(2R_{\lambda\rho}h^{\lambda\rho}+\Box h\right)\right]\\
    (\mathcal{D}^{(4)}[h])_{\mu\nu} =&\frac{1}{32m^2}\left[ \left(3R^2-8R_{\lambda\rho}R^{\lambda\rho}-4\Box R \right)h_{\mu\nu}+\left(16R_{\mu}^{\ \ \lambda}R_{\lambda\nu}-6R_{\mu\nu}R\right.+4\Box R_{\mu\nu}\right. \\
    &\left.+4\nabla_{\mu}\nabla_{\nu}R\right) h-4\left(2R_{\mu\lambda}R_{\nu\rho}-3R_{\mu\nu}R_{\lambda\rho}-R_{\mu\lambda}^{\ \ \ \ \alpha\beta}R_{\nu\alpha\rho\beta}+R_{\mu \ \ \lambda}^{\ \ \alpha \ \ \beta}R_{\nu\beta\rho\alpha}\right.\nonumber\\
    &\left. +6R_{\mu \ \ \nu}^{\ \ \alpha \ \ \beta}R_{\lambda\alpha\rho\beta}-4\nabla_{\rho}\nabla_{\left(\nu\right.}R_{\left.\mu\right)\lambda}+4\nabla_{\rho}\nabla_{\lambda}R_{\mu\nu}\right)h^{\lambda\rho}+g_{\mu\nu}\left(8\nabla_{\alpha}R_{\lambda\rho}\nabla^{\alpha}h^{\lambda\rho}\right.\nonumber\\
    &\left.-3R\Box h +2\Box\Box h +4R^{\lambda\rho}\left(2\nabla_{\lambda}\nabla_{\rho}h-4\nabla_{\alpha}\nabla_{\lambda}h_{\rho}^{\ \ \alpha}+3\Box h_{\lambda\rho}\right) \right)-3R_{\mu\nu}\Box h\nonumber\\
    &-2\left( \ 3 R(2\nabla_{\lambda}\nabla_{\left(\mu\right.}h_{\left.\nu\right)}^{\ \ \lambda}-\nabla_{\mu}\nabla_{\nu}h-\Box h_{\mu\nu})+\nabla^{\lambda}R(\nabla_{\lambda}h_{\mu\nu}-2\nabla_{\left(\mu\right.}h_{\left.\nu\right) \ \lambda})\right.\nonumber\\
    & +2R^{\lambda\rho}\left(3\nabla_{\left(\mu\right.}\nabla_{\left.\nu\right)}h_{\lambda\rho}-8\nabla_{\lambda}\nabla_{\left(\mu\right.}h_{\left.\nu\right)\rho}+4\nabla_{\lambda}\nabla_{\rho}h_{\mu\nu}\right) +2R_{\mu}^{\ \ \lambda}\nabla_{\nu}\nabla_{\lambda}h\nonumber\\
    &+2R_{\nu}^{\ \ \lambda}\nabla_{\mu}\nabla_{\lambda}h-4R_{\mu}^{\ \ \lambda}\nabla_{\rho}\nabla_{\lambda}h_{\nu}^{ \ \ \rho}-4R_{\nu}^{\ \ \lambda}\nabla_{\rho}\nabla_{\lambda}h_{\mu}^{ \ \ \rho}+8\nabla_{\rho}R_{\lambda\left(\mu\right.}\nabla_{\left.\nu\right)}h^{\lambda\rho}\nonumber\\
    &-8\nabla_{\lambda}R_{\rho\mu}\nabla^{\rho}h_{\nu}^{\ \ \lambda }-8\nabla_{\lambda}R_{\rho\nu}\nabla^{\rho}h_{\mu}^{\ \ \lambda }+2\nabla_{\left(\mu\right.}R_{\left.\nu\right)\lambda}\nabla^{\lambda}h +16\nabla_{\lambda}R_{\rho \left(\mu\right.}\nabla^{\lambda}h_{\left.\nu\right)}^{\ \ \rho}\nonumber \\
    &-2\nabla_{\lambda}R_{\mu\nu}\nabla^{\lambda}h+8R_{\mu\lambda\nu\rho}\Box h^{\lambda\rho}-2\nabla_{\left(\mu\right.}R_{\left.\nu\right)\lambda\rho\alpha}\nabla^{\alpha}h^{\lambda\rho}+\nabla_{\lambda}R_{\mu\rho\nu\alpha}\nabla^{\alpha}h^{\lambda\rho}\nonumber\\
    &+\nabla_{\lambda}R_{\nu\rho\mu\alpha}\nabla^{\alpha}h^{\lambda\rho}-2R_{\mu\lambda\rho\alpha}\nabla^{\alpha}\nabla_{\nu}h^{\lambda\rho}-2R_{\nu\lambda\rho\alpha}\nabla^{\alpha}\nabla_{\mu}h^{\lambda\rho}-2R_{\mu\lambda\nu\rho}\nabla_{\lambda}\nabla_{\rho}h\nonumber\\
    &\left.\left.+\nabla_{\lambda}\nabla_{\rho}\nabla_{\mu}\nabla_{\nu}h^{\lambda\rho}+\nabla_{\lambda}\nabla_{\rho}\nabla_{\nu}\nabla_{\mu}h^{\lambda\rho}+2\Box\nabla_{\mu}\nabla_{\nu}h-8\Box\nabla_{\lambda}\nabla_{\left(\mu\right.}h_{\left.\nu\right)}^{\ \ \lambda}+4\Box\Box h_{\mu\nu}\right)\right]\nonumber.
\end{align}

% %\renewcommand{\refname}{...} % For modifying the bibliography heading

% \phantomsection % use it for correct TOC link !!!
% \addtocontents{toc}{\protect\addvspace{4.5pt}}% add vertical space in TOC
% % %\addcontentsline{toc}{section}%{References} % add References to %TOC
% % \bibliographystyle{mybibstyle}
% % \bibliography{bibliografia}
% \providecommand{\href}[2]{#2}\begingroup

\bibliographystyle{JHEP}

\end{document}